\documentclass[psfig,epsfig]{aa}
\usepackage{graphics}
\usepackage{amssymb}
\begin{document}
\thesaurus{ 06() }
\title{VLA Polarimetry of Two Extended Radio Galaxies}
\author{W. Junor\inst{1}, F. Mantovani\inst{2}, 
R. Morganti\inst{2,3} and  L. Padrielli\inst{2} }
\institute{Institute for Astrophysics, University of New Mexico, Albuquerque, 
NM, USA
\and Istituto di Radioastronomia del CNR, Bologna, Italy
\and Australian Telescope National Facility, Epping, NSW 2121,
Australia. }
\offprints{fmantovani@ira.bo.cnr.it}
\date{ Submitted 1996 }
\titlerunning{Polarimetry of Two Extended R.G.}
\maketitle
\markboth{W. Junor et al.}{VLA Polarimetry of Two Extended Radio Galaxies}
\begin{abstract}
Multi-wavelength VLA observations of two extended radio galaxies, $0235-197$
and $1203+043$ are presented.  There is some evidence from earlier
studies that these two sources exhibit low frequency ($<1$\,GHz) variability.  
This work shows that both sources have linear polarizations, if any,
below the detection limits at 320 MHz,
so we cannot explain  the variability as being due to
instrumental polarization effects as has been suggested for 3C159. 
Refractive scintillation may be the cause of the
variability in 0235$-$197.  This would require the existence of a
bright, compact component in one of the hot spots seen in these 
observations.  This is not implausible but the resolution of this
observational program is insufficent to address that question.
The radio source 1203$+$043 lacks any bright compact 
component thereby ruling out a refractive scintillation mechanism for
its variability.  Consequently, it is possible that claims of
variability in this source are spurious.  However, the 320\,MHz VLA 
observations show that 1203$+$043 has an `X'-shaped radio structure.
This is a rare morphology for the brightness distribution of a radio
galaxy; the implications of this are examined.
\end{abstract}
\section{Introduction}
As part of our investigation of steep-spectrum ($\alpha >0.5$, 
S$\propto\nu^{-\alpha}$), low-frequency-variable (LFV;
$\nu <$ 1 GHz) sources, we have made a series of images with 
sub-arcsecond resolutions (Mantovani et al. 1992) of a sample of sources.
These sources were selected from the papers of Cotton (1976), 
McAdam (1980), Spangler \& Cotton (1981), Fanti et al. (1983) 
and Altschuler et al. (1984).  The aim was to detect the 
high-brightness components required by the refractive scintillation 
model for low-frequency variability (Rickett 1986).
Most of the sources in the sample showed compact features (deconvolved 
sizes $<$ 0.15
-- 0.3 arcsec) both in MERLIN, 408 MHz and VLA, A-array, 5 GHz images
(Mantovani et al. 1992).  Further observations with VLBI show these 
features to contain components which are bright and compact enough to
explain the variability at low frequency by propagation 
effects in the interstellar medium; see,  for
example, 3C99, Mantovani et al. (1990a). Sources such as these do not
generally show any variability at high frequency (Padrielli et al.
1987). 
\medskip

However, there are sources like 0621+400 (3C159),
which are variable at low frequencies and not at frequencies
$>2$\,GHz, where the 
compact components are too weak to account for the observed 
variability.  In these cases, it is possible that the 
variability is caused by instrumental effects.
The source 3C159 has been monitored 
for about 10 years at 408 MHz.  This source has a steep radio 
spectrum and an extended double radio structure --- a combination
which is very unusual for a variable radio source. Browne et al.\
(1985) have suggested that the variations, which seem to show an annual
cycle, may not be intrinsic but could arise from the combined
effects of strong source linear polarization and ionospheric Faraday
rotation.

Ionospheric Faraday rotation can easily reach 7--8 rad m$^{-2}$
(Sakurai \& Spangler 1994). With the plausible estimate of 5 rad m$^{-2}$
as an expectable difference in the ionospheric RM, one finds that the
position angle difference at 408 MHz is 2.7 radians; enough to produce
the effect being discussed.

MERLIN observations at 408 MHz (Cerchiara et al. 1994)
show that 3C159 is highly polarized ($\sim$10$\%$).
The plane of polarization of the source emission could be rotated by
changes in ionospheric Faraday rotation relative to the 
linearly polarized E-W arm of the Northern Cross Bologna telescope
used for the monitoring program.  A source with a linear polarization 
$>$6$\%$ could exhibit apparent variations of roughly the observed
size if the ionospheric Faraday rotation changed by $\sim$90$^o$ between 
observations.  The 3C159 observations were made at transit during the
day in summer and the night in winter and so they were accompanied by annual 
changes in the ionospheric electron content.

The purported variability measured in the two extended radio sources 
0235$-$197 and 1203$+$043 may have originated in a similar manner to 
that in 3C159.  They were monitored at 408 MHz 
with a similar instrument, the Molonglo Cross.  With
a peak-to-peak fractional variability of $\sim$10\%, 
they were classified as `probably variable' by McAdam  (1980).

Consequently, although it was expected that most of the sources 
belonging to our sample of steep-spectrum, low-frequency-variable 
sources would be core-dominated sources, it is likely that the sample
has been contaminated by lobe-dominated, strongly-linearly-polarized, 
extended sources.
\hfill\break

In order to test if ionospheric Faraday rotation is the cause of the
apparent variability of 0235$-$197 and 1203$+$043 we have investigated
the linear
polarizations of these sources at 320\,MHz with the VLA in the `A'
configuration and with the already available 5\,GHz VLA C-array data. 
Both sources were also observed in the X (8.4~GHz) and U (15~GHz) bands.  These
observations allowed high resolution images of the 'hot spot' regions
to be made.  The
images were combined with available, high-resolution, C band observations
to produce rotation measures (RMs) for the outer parts of the sources. 

\section {VLA Observations}

VLA (Thompson et al.\ 1980) observing dates and observational
parameters are summarized in Table~1.  The high resolution (A-array) 
Total Intensity images at 5 GHz for 0235$-$197 and 1203$+$043 were
presented in Mantovani et al.\ (1992) while the lower resolution
C-array, total intensity map for 0235$-$197 has been published in
Morganti, Killeen \& Tadhunter (1993).  However, because the
polarization information was never presented in the earlier papers, we 
have summarized in the tables all the observational details and 
the derived parameters from those observations.
%
% Table 1 in here!!!
%
%
\begin{table*} 
%\begin{flushleft}
%\small
\scriptsize
%\footnotesize
\begin{center}
\caption{VLA observing dates}
\medskip
\begin{tabular}{|l|r|c|c|c|} 
\hline  
Source& Band  & Array & Duration & Observing  \\
      & MHz   &       & minutes  & Date       \\
\hline \hline
0235$-$197 &    320 & A & 87 & 22 Dec 1992 \\
           &   4885 & A & 30 & 30 May 1986 \\
           &   4885 & C & 15 & 27 May 1989 \\
           &   8440 & A & 49 & 10 Sep 1990 \\
           &  14940 & A & 51 & 10 Sep 1990 \\
1203$+$043 &    320 & A & 67 & 09 Dec 1992 \\
           &   4885 & A & 39 & 30 May 1986 \\
           &   8440 & A & 28 & 10 Sep 1990 \\
           &  14940 & A & 32 & 10 Sep 1990 \\
\hline
\end{tabular}
\end{center}
\end{table*}
Because of narrowband interference, the P (320\,MHz) band data were
taken in spectral line mode.  The data were edited to remove channels
with interference.  Bandpass corrections were determined from the
calibrator source 3C286 (1331+305) and applied to the spectral line
database.  A new ``Channel 0'' database was then constructed and the
data were calibrated for total intensity and polarization in the
standard fashion (see, for example, Perley, Schwab \& Bridle 1989).
Instrumental polarization calibration was done using the 
calibration sources 3C48 (0134$+$329), 3C138 (0521$+$166) and 
3C286 (1331+305) to get sufficient parallactic coverage across both
days during which the target sources were observed. We assume that the
instrument is stable between observing epochs.

The parallel hand (RR, LL) data were self-calibrated and imaged in the
normal iterative manner.   The complex gain corrections derived from
self-calibration were also applied to the cross-hand (RL or LR)
fringes.  In turn, images in Stokes parameters I, Q, U and V were
produced.  Maps of the polarized flux density $P=(Q^{2}+U^{2})^{1/2}$
and position angle $\chi=0.5\times \tan^{-1}(U/Q)$, were then generated
from the Q and U images.  The Stokes V images were used to test the integrity
of the calibration and self-calibration procedures and to diagnose
problems due to interference.

At low frequencies, the primary beam of the antennas contains many 
background sources.  In order to image the target sources of interest,
it was necessary to image some of these background sources. 
% XXX --- We did not do this the second time around.
% We
% first made a low resolution image of the whole 320\,MHz primary beam
% ($\approx$156\,arcmin diameter) in order to
% identify the brightest confusing sources. 
The brightest ($>20$\,mJy at L~band) background sources were identified 
from the NRAO VLA Sky Survey (NVSS) %\citep{cond:98}.  
% Oh, we haven't used the power of \theblibliography environment.
(Condon et al.\ 1998). This threshold 
is somewhat arbitrary but gives a reasonable number of secondary fields 
to image at 320\,MHz.  We imaged a total of 
26 fields for 0235--197 and 23 fields for 1203$+$043 in the {\it AIPS} 
mapping program, IMAGR --- one field containing the program source and 
the others on the
brightest background sources.  We were able to account for
substantially all of the flux density seen on the shortest baselines
in both cases and to obtain satisfactory convergence in the iterative
self-calibration and imaging loop.  The final images in all Stokes
parameters (I, Q, U \& V) of the target sources contained no obvious
artefacts due to sidelobes from nearby confusing sources.  
The linear polarizations of the background sources have been checked
in order to test for beam squint between R and L beams. 
The r.m.s.\ noises in the final images are within a factor of 3 of 
the expected thermal noises; this is entirely consistent with other 
observing programs at 320\,MHz. Together, these suggest that the images 
of the target sources are not  greatly affected by confusion. 

The low declination source 0235--197  was observed for only $\approx2$ 
hours at transit; consequently, some of the data were corrupted by 
cross-talk between antennas.  These data, mostly on baselines with 
immediately adjacent antennas, were excised from the database during 
the iterative self-calibartion and imaging process.  

The C (5~GHz), X (8.4~GHz) and U (15~GHz) band data were calibrated
in the standard way using VLA calibrators and AIPS procedures.
Polarimetric images were generated in a manner similar to that for the
P band data. 
  
\section {Sources structure and Polarimetry}

\subsection {Observational Parameters}

The derived parameters for the low resolution observations at 320\,MHz
and at 5\,GHz (VLA C-array) are listed in Table~2.  Maps at higher
resolution have been obtained at 8.4 and 15 GHz with the VLA in the A
configuration.  Values derived from the maps are listed in Table~3. 
Comments on the sources structure will be given in Section~4.

The contents of Tables 2 and 3 are: column 1 $-$
source name; column 2 $-$ the observing frequency in MHz;
columns 3 to 5 $-$ major axis, minor axis (both in arcsec) and the PA
in degrees of the restoring beam major axis; column 6 $-$ the rms noise in 
the total intensity map far from the source of emission; 
column 7 $-$ the rms noise
$\sqrt{\sigma^{2}_Q+\sigma^{2}_U}$, where $\sigma{_Q}$ and $\sigma{_U}$ 
are the
rms noises on the blank sky in the distributions of the Stokes parameters
Q and U; column 8 $-$ component label; columns 9 and 10$-$ RA and Dec. of the
component peak; column 11 $-$ peak flux density (mJy) of the component;
column 12 $-$ total flux density (mJy) of the component.

\medskip

In Tables 4 \& 5, we give the measured position angle (PA) in
degrees of the electric field vector at the peak of polarized
emission ($\pm$1 rms error calculated from the distribution of PAs
found in a small box around the peak of polarized emission); the 
Rotation Measure 
(RM$=\Delta\phi(\lambda)\pm n\pi/\Delta(\lambda^2)$ in $rad\,m^{-2}$
where $\phi$($\lambda$) is the PA at wavelength $\lambda$ and $n$ an integer;
when three frequencies are available, as for some of the components in
Tab.\,4, the ambiguity inplied by the integer $n$ can be resolved); the RM, 
corrected for the redshift; the percentage polarization; the depolarization
index, defined as the ratio of the fractional polarization at longer
wavelength to the fractional polarization at shorter wavelength;
from the high and low resolution observations. 

In order to compare the 5\,GHz (C-Array) and 320\,MHz images of
0235$-$197, we have produced 5\,GHz maps (I,Q,U) at the resolution
of the 320\,MHz maps.  This was done by restoring the 5\,GHz images with
the appropriate Gaussian beam during imaging.  The polarization
parameters derived from those images are reported in Table~5.
%
% Table 2 here!!!
%
%
\begin{table*} 
%\begin{flushleft}
%\small
\scriptsize
%\footnotesize
\begin{center}
\caption{Low resolution observational parameters and observed properties.}
\medskip
\begin{tabular}{lrccrccrrrrr} 
\hline 
Source& Obs.&         &Beam     & PA   &$\sigma_t$&$\sigma_p$&C&   
R.A.(J2000)&  Dec.(J2000)&  Flux &Dens.   \\
      &$\nu$& maj.    &   min.  &      &          &          &     &   
 &    & peak& total  \\
      & MHz &$\arcsec$&$\arcsec$&$\degr$& mJy/b    &  mJy/b   &     & h~~~m~~~s~~~~&
$\degr$~~~$\arcmin$~~~$\arcsec~~~~$&mJy/b&mJy   \\
\hline \hline
0235$-$197& 320& 7.3 & 4.3    & --17   & 4.1     & 4.5     & E &02~37~44.5
& --19~32~30 & 3430   & 8530   \\
          &     &     &         &      &          &        & C &~~~~~~42.9
& ~~~~~~~~25 & 1068 & 3190  \\
          &     &     &         &      &          &        & W &~~~~~~41.9
& ~~~~~~~~28 & 1117 & 2610  \\
% 6cm da sistemare
      & 4885    & 4.5 & 2.2    &  -70  & 0.3     & 0.1     & E &~~~~~~44.6
& ~~~~~~~~36 &281.6&854.0 \\
          &     &     &         &      &          &          & C &~~~~~~43.0
& ~~~~~~~~31 & 35.2&145.3 \\
          &     &     &         &      &          &          & W &~~~~~~42.0
& ~~~~~~~~34 & 70.4&266.5 \\

1203$+$043& 320 & 5.6 & 5.2    &--2  & 1.1     & 1.3     &  N  &12~06~20.4
& ~04~06~21 & 796   &  -- \\ 
          &     &     &         &      &          &          &  S  &~~~~~~19.6
& ~05 &   579   &  --        \\
\hline
\end{tabular}
\end{center}
\end{table*}
%
% Table 3 here!!!
%
%
\begin{table*} 
%\begin{flushleft}
%\small
\scriptsize
%\footnotesize
\begin{center}
\caption{High resolution observational parameters and observed properties.}
\medskip
\begin{tabular}{lrccrcclrrrr} 
\hline 
Source& Obs.&         &Beam     & PA   &$\sigma_t$&$\sigma_p$&C&   
R.A.(B1950)&  Dec.(B1950)&  Flux &Dens.   \\
      &$\nu$& maj.    &   min.  &      &          &          &     &   
 &    & peak& total  \\
      & MHz &$\arcsec$&$\arcsec$&$\degr$& mJy/b    &  mJy/b   &     & h~~~m~~~s~~~~&
$\degr$~~~$\arcmin$~~~$\arcsec~~~~$&mJy/b&mJy   \\
\hline \hline
0235$-$197& 4860& 0.55 & 0.37   & 18   & 0.05     & 0.14  &$E_1$&02~35~26.06
& --19~45~34.22 & 74.6   & 128.6 \\
          &     &     &         &      &          &          &$E_2$&~~~~~~26.02
& ~~~~~~~~34.60 & 33.6   &  58.8 \\
          &     &     &         &      &          &          & W   &~~~~~~23.04
& ~~~~~~~~30.89 &  4.0   &  14.5 \\

          & 8440&0.36 & 0.26    &  16  & 0.07     & 0.03     &$E_1$&~~~~~~26.06
& ~~~~~~~~34.17 & 44.0& 109.7 \\
          &     &     &         &      &          &          &$E_2$&~~~~~~26.01
& ~~~~~~~~34.63 & 16.4&  51.4 \\
          &     &     &         &      &          &          &  W  &~~~~~~23.14
& ~~~~~~~~30.90 &  1.5&   7.8 \\

          &14940&0.19 & 0.14    &  24  & 0.15     & 0.10    &$E_{1a}$&~~~~~~26.07
& ~~~~~~~~34.13 & 15.7& 26.0  \\
          &      &    &         &      &          &         &$E_{1b}$&~~~~~~26.07
& ~~~~~~~~34.37 & 11.0& 18.0  \\
          &      &    &         &      &          &          &$E_2$&~~~~~~26.02
& ~~~~~~~~34.60 &  2.2& 11.0  \\

1203$+$043& 4860&0.41 & 0.40    &--41  & 0.06     & 0.07     &  N  &12~03~46.58
& ~04~22~58.50 & 1.4  & 49.4  \\
          &     &     &         &      &          &          &  C  &~~~~~~46.48
& ~52.90 &   6.2   &  6.7        \\
          &     &     &         &      &          &          &$J_1$&~~~~~~46.23
& ~47.46 &   24.0   &  55.4       \\
          &     &     &         &      &          &          &$J_2$&~~~~~~45.94
& ~43.90 &   6.6   &  44.9        \\
          &     &     &         &      &          &          &  S  &~~~~~~
& ~       &  0.8   &  33.3         \\

          & 8440&0.30 & 0.29    &  48  & 0.06     & 0.07     &  N  &12~03~46.71
& ~04~23~01.65 & 0.78   & 325  \\
          &     &     &         &      &          &          &  C  &~~~~~~46.47
& ~52.67 &   8.7   &  8.7        \\
          &     &     &         &      &          &          &$J_1$&~~~~~~46.22
& ~47.65 &   14.8   &  30.2       \\
          &     &     &         &      &          &          &$J_2$&~~~~~~45.94
& ~44.12 &   3.0   &  15.2        \\
          &     &     &         &      &          &          &  S  &~~~~~~
& ~       &  0.5   &  4.0         \\

          &14940&0.18 & 0.16    &  50  & 0.06     & 0.10     &  C  &~~~~~~46.47
& ~52.66 & 14.5 & 14.9  \\
          &     &     &         &      &          &          &$J_1$&~~~~~~46.22
& ~47.63 &  4.3   &  8.2 \\
\hline
\end{tabular}
\end{center}
\end{table*}
\section {Notes on individual sources}
\subsection {0235$-$197}
The 5 GHz images presented by Mantovani et al. (1992) and Morganti,
Killeen \& Tadhunter (1993), show the classical, double structure
typical of  powerful radio galaxies.  The source is associated with a
galaxy at $z=0.620$ (Tadhunter et al. 1993). There are no radio
components above the detection limits ($\leq$0.2 mJy at 5 and 8.5 GHz;
$\leq$0.5 mJy at 15 GHz) inside the error box of the optical position.
Adopting the optical position as a reference, 0235$-$197 looks rather
symmetric with a ratio of $\sim$0.8 between the full length of the two
lobes, with the western  being the longer.

0235$-$197 appears to be dominated by the outer lobes at frequencies
$> 5$ GHz. The most interesting feature is the bright hot spot at the
far end of the Eastern lobe. At both 5 and 8.4 GHz (Figs.\,1 to 3), the hot 
spot has a double structure with individual components labelled {\it $E_1$}
and {\it $E_2$}. Component {\it $E_1$} also appears double when
observed with higher resolution at 15\,GHz (Fig.\,4). The images have
been convolved to 0.4 arcsecond resolution in order to calculate the hot spot 
spectral indices. Both components {\it $E_1$} and {\it $E_2$} show a
large steepening in spectral index, $\alpha$ (S$\propto\nu^{-\alpha}$),
between 5--8.4\,GHz and 8.4--15\,GHz. We find values of $\alpha = 0.5$ 
and $\alpha = 1.4$ respectively for {\it $E_1$} and $\alpha = 0.68$ and 
$\alpha = 2.5$ respectively for {\it $E_2$}.  The front shock of the
lobe {\it $W_1$} at the opposite side is resolved in all of the images
and can hardly be defined as a `hot spot'. The spectral index of the
bright part at the far end is also steep ($\alpha = 1.2$) in the 
range 5--8.4 GHz and it steepens to $\alpha\geq 2$ in the range 8.4-15 GHz
(since the 15 GHz flux density estimate is an upper limit).
Note that the observations at 15 GHz have lower sensitivity to
diffuse, extended  emission.
%
%              fig.1
\begin{figure}
\resizebox{\hsize}{!}{\includegraphics{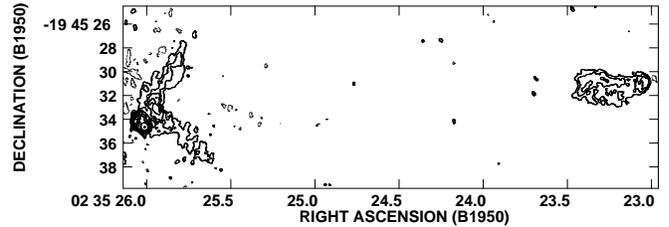}}
\caption[]{VLA image of 0235$-$197 at 8.4 GHz. Contours 
are at $-$0.3, 0.3, 0.6, 1, 2, 4, 8, 16, 32, 64
mJy\,beam$^{-1}$. The peak flux density is 45.4 mJy\,beam$^{-1}$.}
\end{figure}
%
%              fig.2
\begin{figure}
\resizebox{\hsize}{!}{\includegraphics{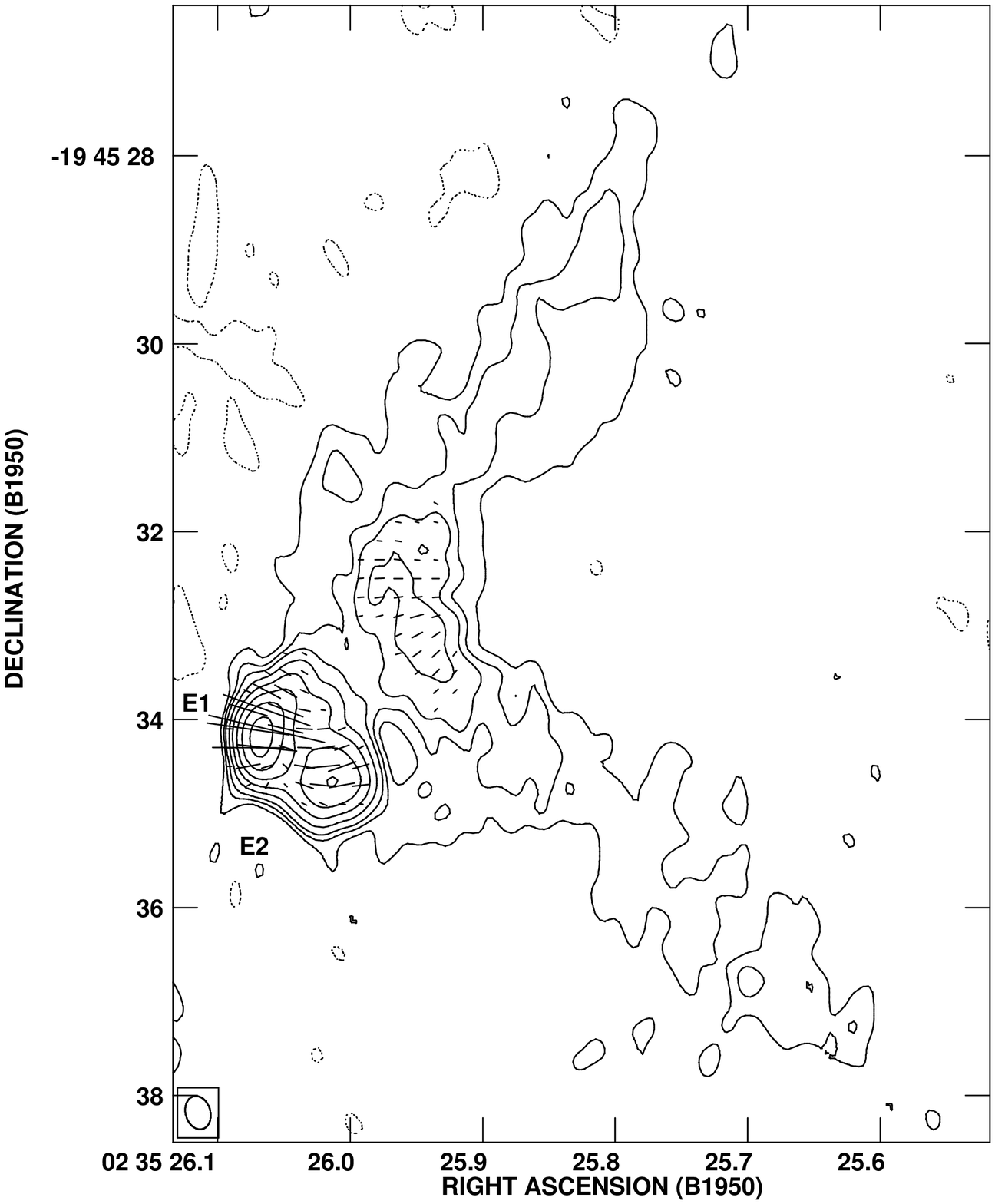}}
\caption[]{VLA image of the East lobe of 0235$-$197 at 8.4 GHz. Contours 
are at $-$0.3, 0.3, 0.6, 1, 2, 4, 8, 16, 32, 64
mJy\,beam$^{-1}$. A vector length of 1$\arcsec=$ 10 mJy\,beam$^{-1}$.}
\end{figure}
%
%              fig.3
\begin{figure}
\resizebox{\hsize}{!}{\includegraphics{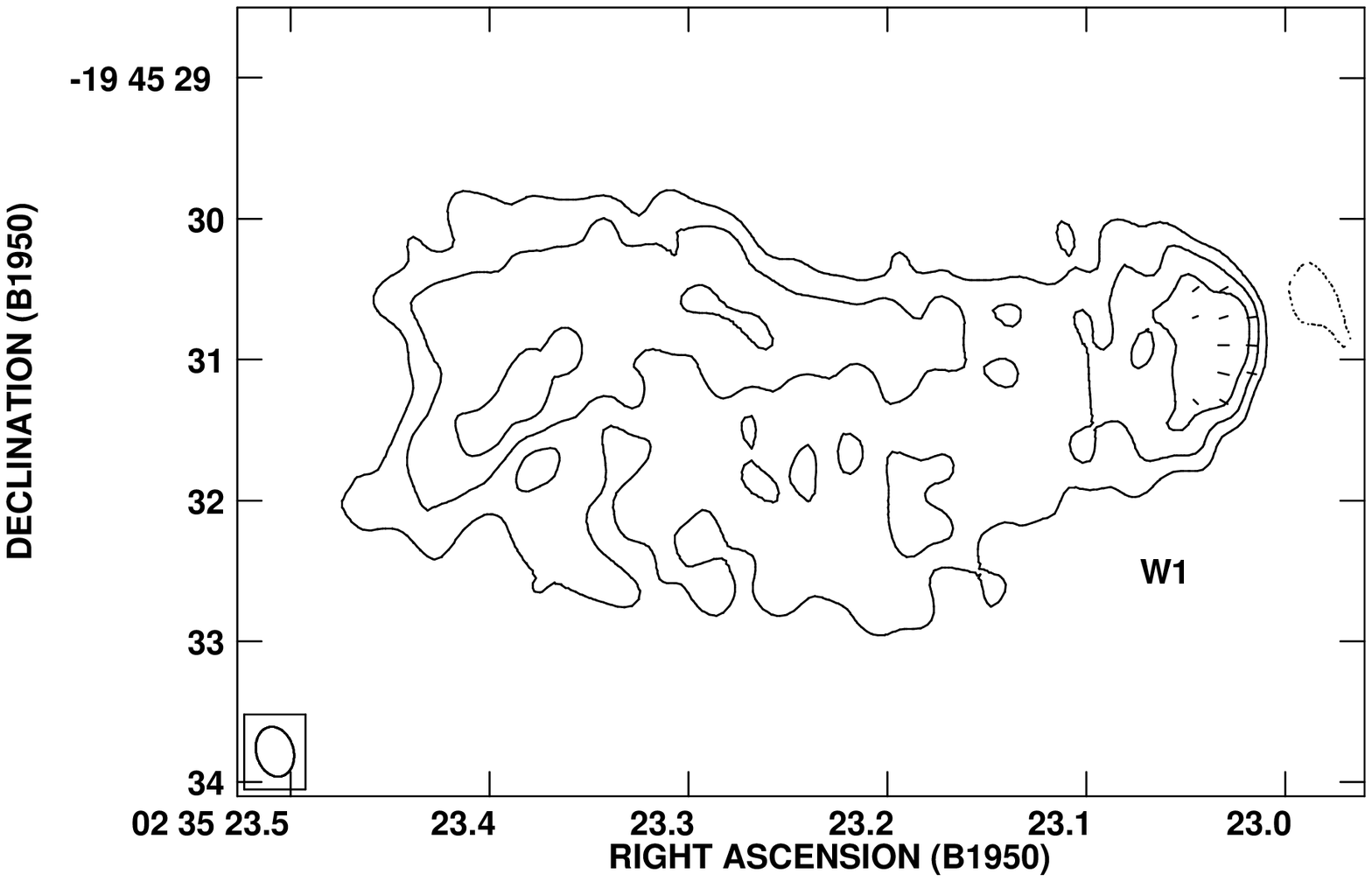}}
\caption[]{VLA image of the West lobe of 0235$-$197 at 8.4 GHz. Contours 
are at $-$0.3, 0.3, 0.6, 1, 2, 4, 8, 16, 32, 64
mJy\,beam$^{-1}$. A vector length of 1$\arcsec=$ 10 mJy\,beam$^{-1}$.}
\end{figure}
%
%              fig.4
\begin{figure}
\resizebox{\hsize}{!}{\includegraphics{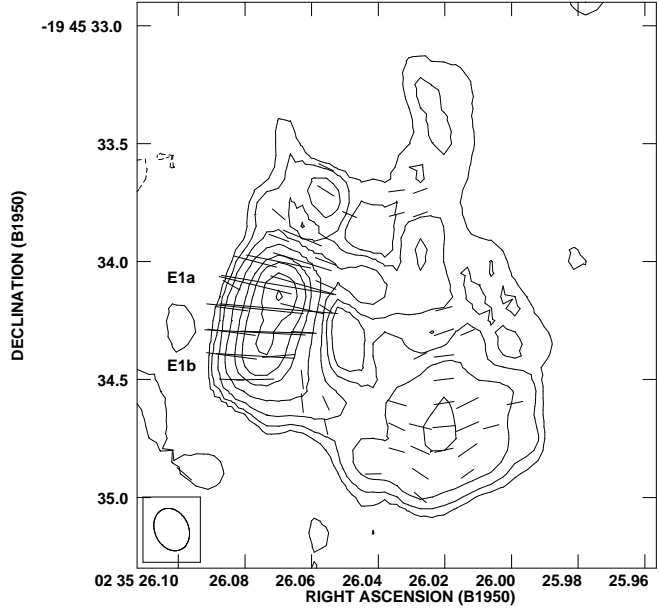}}
\caption[]{VLA image of the East lobe of 0235$-$197 at 15 GHz. Contours 
are at $-$0.4, 0.4, 0.6, 1, 2, 4, 6, 10, 15, 20 mJy\,beam$^{-1}$. 
The peak flux density is 15.7 mJy\,beam$^{-1}$. 
A vector length of 1$\arcsec=$ 6.7 mJy\,beam$^{-1}$.}
\end{figure}
%
%              fig.5
\begin{figure}
\resizebox{\hsize}{!}{\includegraphics{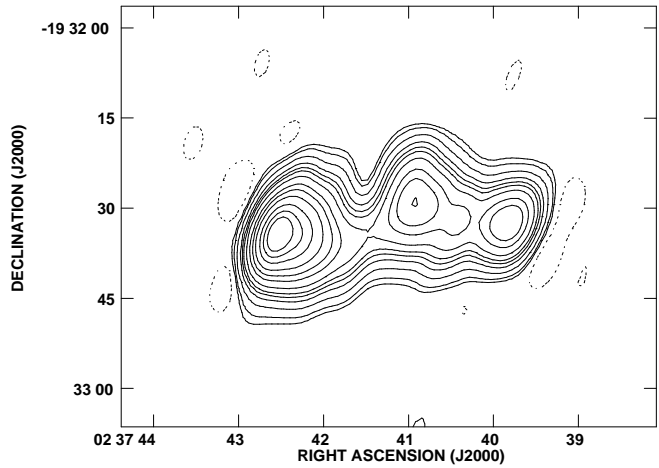}}
\caption[]{VLA image of 0235$-$197 at 320 MHz. Contours 
are at $-$5, 5, 10, 20, 50, 100, 150, 200, 300, 500, 700,
1000, 1500, 2000, 2500 mJy\,beam$^{-1}$. 
The peak flux density is 3430 mJy\,beam$^{-1}$.} 
\end{figure}
%
%              fig.6
\begin{figure}
\resizebox{\hsize}{!}{\rotatebox{-90}{\includegraphics{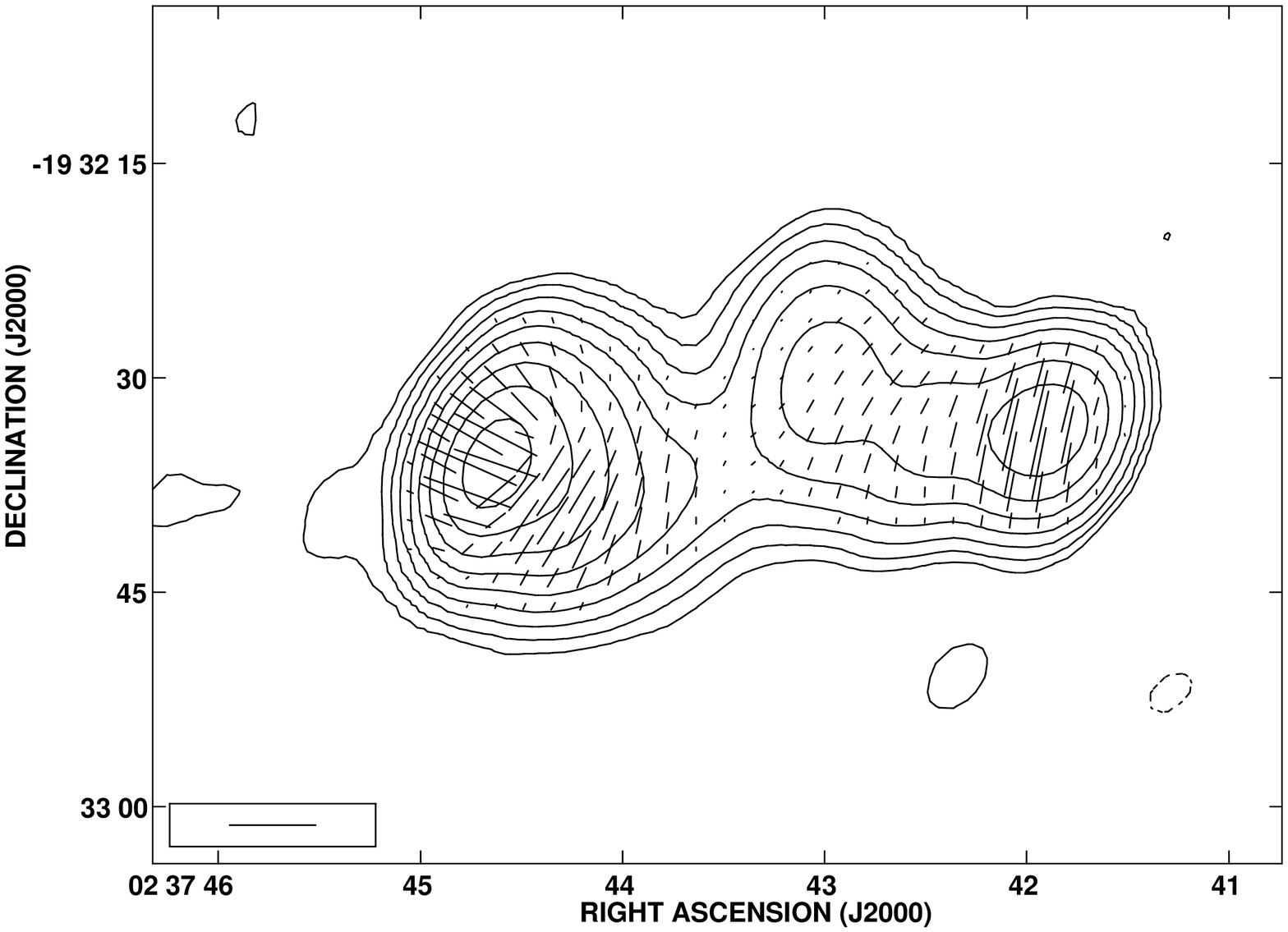}}}
\caption[]{VLA image of 0235$-$197 at 5 GHz. Contours 
are at 1.1 $\times$ ($-$1, 1, 2, 4, 8, 16, 32, 64, 128, 256 mJy\,beam$^{-1}$. 
The peak flux density is 419.1 mJy\,beam$^{-1}$. 
A vector length of 1$\arcsec=$ 5 mJy\,beam$^{-1}$.}
\end{figure}
All of the hot spots are highly polarized, with little depolarization
and Faraday rotation.  The magnetic field is parallel to the
front shock and rather ordered in the lobe regions with weak diffuse
emission. (The electric vector is shown in all of the images shown in 
the paper). At 320 MHz (Fig.\,5), the polarized emission, if any,
is below the detection limit of our observations.
Polarized emission is detected over all of the source 
in the low resolution observations at 5 GHz (Fig.\,6). The magnetic
field is again ordered and, generally speaking, parallel to the source
major axis, apart from the hot spot area, where the magnetic field is
parallel to the front shock.  There are two main regions of polarized
emission in the $E$ lobe. The mean PA given in Table 5 should be treated
with care because the position angles of the polarization 
vectors in the lobe actually vary greatly.  The depolarization
between 5 GHz and 320 MHz is very high (DP$>$0.02). Statistically, in these
classical sources, the lobe nearest to the nucleus usually shows a
steeper spectral index. Here we find similar values
($\alpha=$0.84) for the two lobes of 0235$-$197.   There are
indications of Faraday rotation in the hot spots from the high 
resolution maps. 
\subsection {1203$+$043}
The images at 8.4 and 15 GHz (Figs.\,7 to 8 \& 9 respectively) do not add
much new information about the overall source structure as derived
from the 5\,GHz image of Mantovani et al.\ (1992).  This earlier image
showed a long bent jet. The components found along the jet, labelled
{\it $J_1$} and {\it $J_2$}, are rather polarized.  The emission from
{\it $J_1$} has a small Faraday rotation and depolarization between
6\,cm  and 4\,cm of 17 $rad\,m^{-2}$ and 0.74 respectively.  

However, the new observations have allowed us to identify component
{\it C} with the core of the radio source.   This component has an
inverted spectrum which peaks at frequencies $\geq$ 15 GHz. 
1203$+$043 therefore has an asymmetric structure, with a long bent jet
pointing South which fades slowly and with a weak lobe of emission to
the north where there is marginal evidence of an hot spot.
The radio position of the core of 1203$+$043 does not coincide within the
errors to any optical counterpart on the Palomar Sky Survey prints.
%
%              fig.7
\begin{figure}
\resizebox{\hsize}{!}{\includegraphics{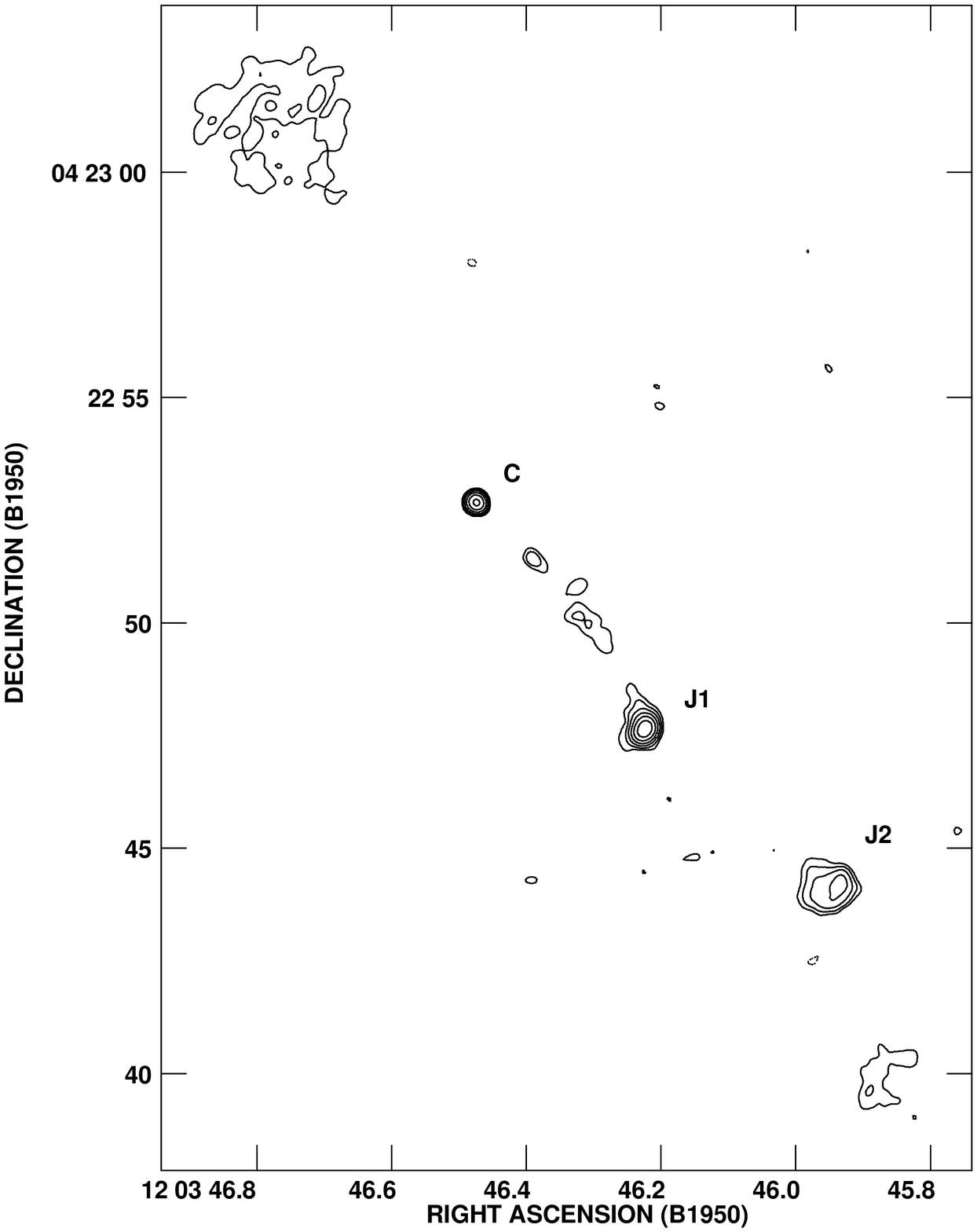}}
\caption[]{VLA image of 1203$+$043 at 8.4 GHz. 
Contours 
are at $-$0.25, 0.25, 0.5, 1, 2, 4, 8, 16, 32, 64, 128 mJy\,beam$^{-1}$.
The peak flux density is 14.8 mJy\,beam$^{-1}$.} 
\end{figure}
%
%              fig.8
\begin{figure}
\resizebox{\hsize}{!}{\includegraphics{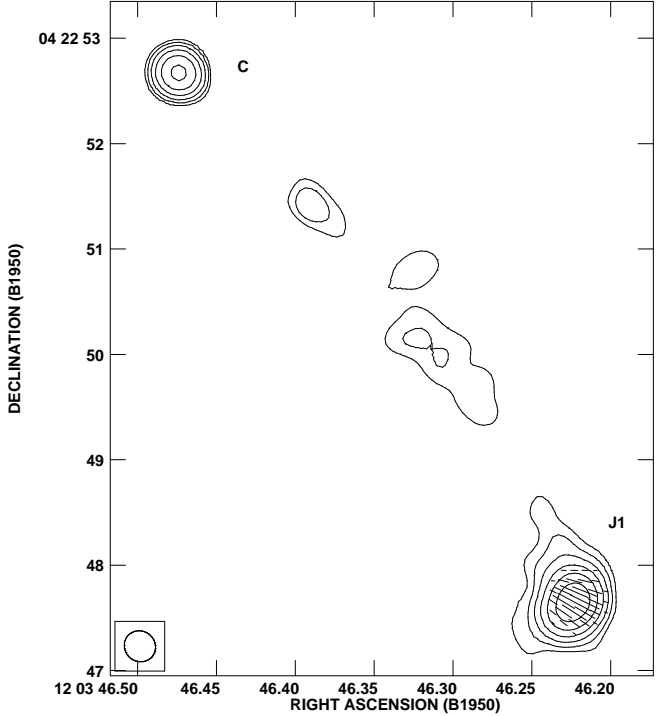}}
\caption[]{VLA image of the central region of 1203$+$043 at 8.4 GHz. 
Contours 
are at $-$0.25, 0.25, 0.5, 1, 2, 4, 8, 16, 32, 64, 128 mJy\,beam$^{-1}$.
The peak flux density is 14.8 mJy\,beam$^{-1}$. 
A vector length of 1$\arcsec=$ 10 mJy\,beam$^{-1}$.}
\end{figure}
%
%              fig.9
\begin{figure}
\resizebox{\hsize}{!}{\includegraphics{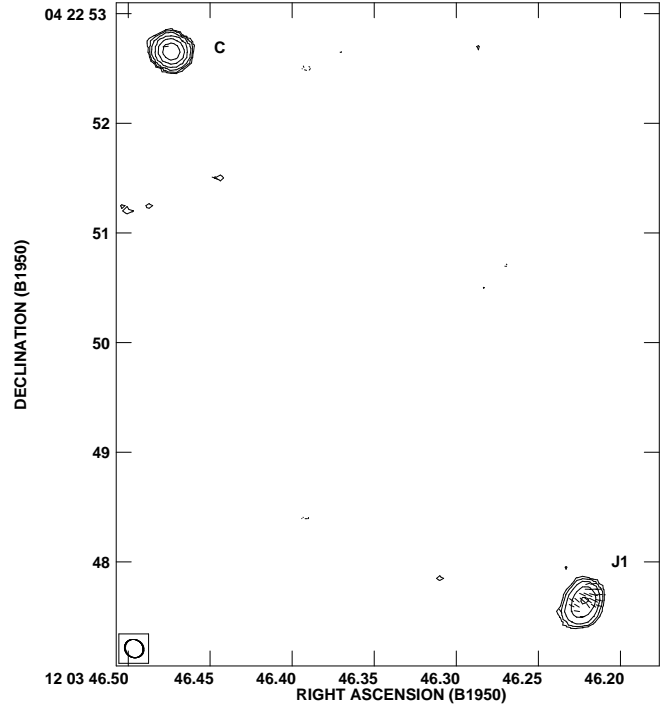}}
\caption[]{VLA image of 1203$+$043 at 15 GHz. 
Contours 
are at $-$0.4, 0.4, 0.6, 1, 2, 4, 8, 16, 32 mJy\,beam$^{-1}$.
The peak flux density is 14.5 mJy\,beam$^{-1}$. 
A vector length of 1$\arcsec=$ 10 mJy\,beam$^{-1}$.}
\end{figure}
%
%              fig.10
\begin{figure}
\resizebox{\hsize}{!}{\includegraphics{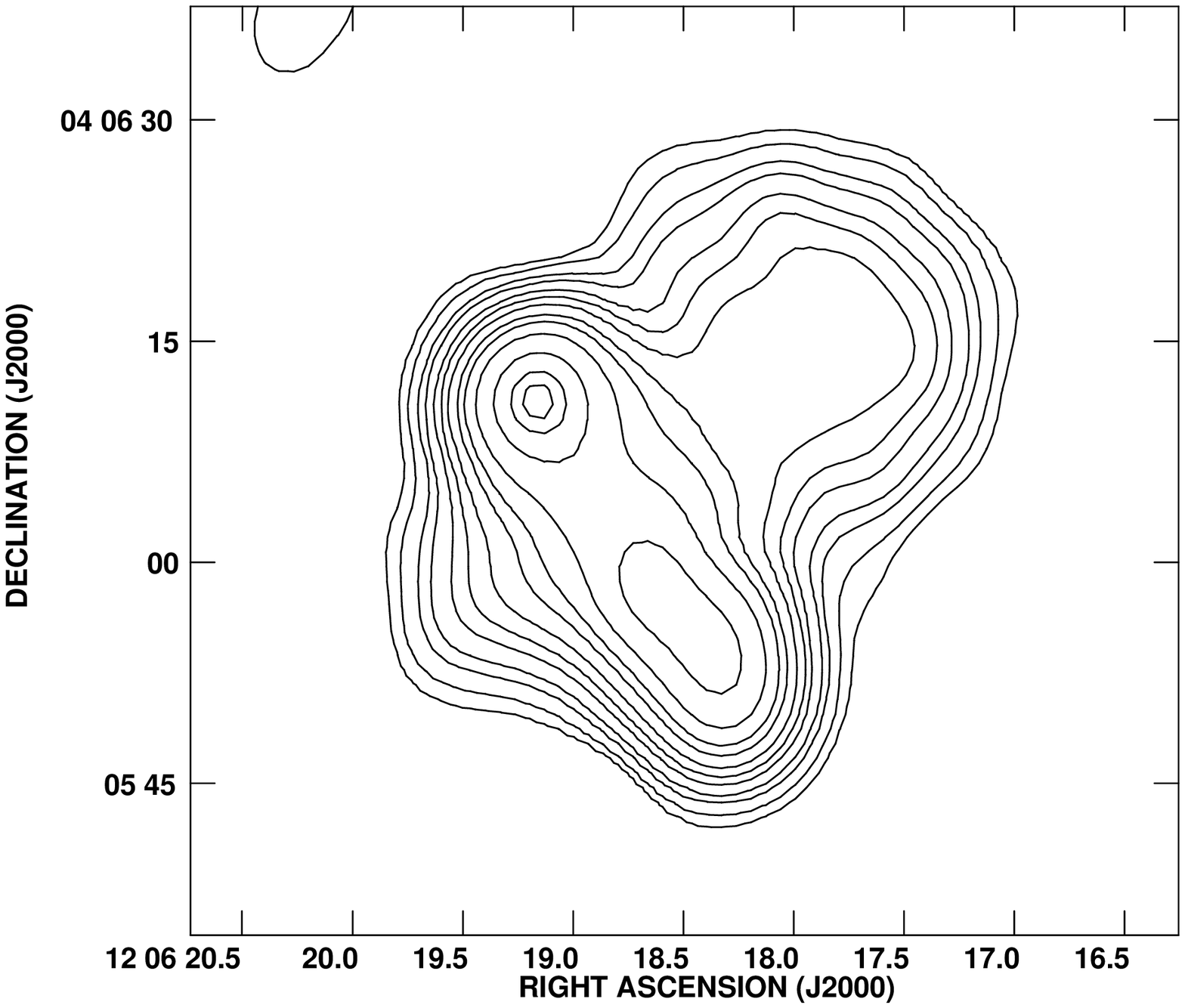}}
\caption[]{VLA image of 1203$+$043 at 320 MHz. Contours 
are at $-$5, 5, 10, 20, 30, 50, 70, 100, 150, 200, 300, 500, 700,
800 mJy\,beam$^{-1}$. 
The peak flux density is 796 mJy\,beam$^{-1}$.}
\end{figure}
Much more interesting is the structure found 
at 320 MHz (Fig.\,10). Together with the North-South structure which
dominates at higher frequencies and which appears here as a ridge of
emission, there is a region of diffuse emission with its major axis
{\it perpendicular} to the main ridge.  This new feature is about 50
arcseconds in extent and is comparable in width with the main
North-South ridge.   
%
%\footnote{The fields of view of the VLA 320~MHz observations contain many
%background sources. The list of sources with derived observational
%parameters can be found on the WEB ... }
%
%\section {Sources in the field at 320 MHz of 0235$-$197 and 1203$+$043}
%
%The fields of view of the VLA 320~MHz observations contain many
%background sources.   Those
%sources are listed in Table 6 together with derived observational 
%parameters.
%The contents of Table 6 are: column 1 $-$
%source name; column 2 $-$ component label;
%columns 3 and 4$-$ RA and Dec. of the component peak;
%columns 5 to 7 $-$ deconvolved size of the component: major axis, minor axis 
%(both in arcsec) and the PA in degrees of the major axis; 
%column 8 $-$ peak flux density (mJy) of the component;
%column 9 $-$ total flux density (mJy) of the component.
%The peak flux density and the total flux density were corrected for the
%primary beam attenuation.
%
% Table 4
%
\begin{table*} 
\begin{flushleft}
\caption{Polarization parameters from high resolution observations.}
\medskip
\begin{tabular}{lccrrrrrrrrrr} 
\hline  
Source&{\it z}& C   &   & PA &     &  RM& RM$\times$&    &$\%$Pol&   & DP& DP  \\
      &       &   &6cm& 4cm& 2cm &      &(1$+${\it z})$^2$& 6cm& 4cm   &2cm&6-4& 4-2 \\
\hline \hline

0235$-$197&0.62& $E_1$ & 68$\pm$~6 & 77$\pm$~2 & 84$\pm$~1 &  87  & 228
&22.0&23.6&20.2 & 0.93 & 1.2  \\

          &    & $E_2$ &-80$\pm$~7 &-84$\pm$~2 & 82$\pm$~2 &  -87  & -228
&13.4&17.8& 8.4 & 0.75 & 2.1  \\

          &    & $W_1$ & 78$\pm$~4 & 83$\pm$~2 &  -        &   35  &   92
&22.8&19.5& -   & 1.20  & -    \\

1203$+$043&    & $J_1$ & 68$\pm$~3 & 62$\pm$~1 & 66$\pm$~6 &   17  & 
&15.0&20.2& 5.5 & 0.74 & 3.7  \\

          &    & $J_2$ &-68$\pm$~8 &-49$\pm$~1 &  -        &  140  & 
&15.8&10.9& -  &  1.45 &  -  \\
\hline
\end{tabular}
\end{flushleft}  
\end{table*}
%
% Table 5
%
\begin{table*} 
\begin{flushleft}
\caption{Polarization parameters from low resolution observations of
         0235$-$197 at 6\,cm.}
\medskip
\begin{tabular}{lcclr} 
\hline  
Source    & C   & PA         & $\%$Pol  \\

\hline \hline

0235$-$197& $E$ & -16$\pm$18 &  10.5      \\
          & $C$ & -85$\pm$1  &   9.1     \\
          & $W$ & -18$\pm$5  &  20.6     \\
\hline
\end{tabular}
\end{flushleft}  
\end{table*}
\section {Discussion}
Due to the non-detection of polarized emission at 320\,MHz in both 0235$-$197 
and 1203$+$043, we cannot explain the low frequency variability observed
with the Molonglo Cross (McAdam 1980) in terms of ionospheric Faraday
rotation as in the case of 3C159. 

Can refractive scintillation (Rickett, 1986) explain the variability
of 0235$-$197?  
The refractive scintillation models assume a supposedly-variable radio
source to have most of its flux density in a single compact
component.  The degree of variability is a function of the
characterization of the interstellar medium (itself a function of 
galactic coordinates) and source size.
0235$-$197 is at galactic latitude $|b|=65^{\circ}$ and
was reported to vary (rms variability $\sim 0.2$\, Jy)
on a time scale of the order of 1 year (McAdam, 1980). 

%Considering the refractive theory applied to consuetudinal model for 
In the usual refractive model of 
interstellar turbulence (Mantovani et al. 1990b, Spangler et al. 1993, 
Spangler et al. 1994, Bondi et al. 1994), the relevant parameter for 
the observed scintillation index is $\theta_{eff}^{7/6} \sqrt{\sin b/I}$,
where $\theta_{eff}=\theta_{\mathrm{FWHM}}/2.35$ and $I$  is the parameter indicator
of the source structure ($=1$ for a gaussian structure).
In such a model a source of ($\approx 3$\,Jy) with a rms variability of 0.2\,Jy
and corresponding scintillation index of $\sim 0.06$ should have an angular
diameter in the range 10--20 mas (see Spangler et al. 1993 for details).

The $E_{1a}$ hot spot has a spectrum which, extrapolated towards lower
frequencies, gives a mean flux density of $\sim$3\, Jy at 408 MHz.
This is comparable to the peak flux density found at 320 MHz. 
The crucial parameter is, however, the angular size of the hot spot.
The deconvolved size found for the $E_{1a}$ hot
spot at 15 GHz is $<$0.2\, arcsec.  Even if in principle there is not
contradiction, the scintillation theory requires an angular size for the
hot spot in 0235$-$197 that is a factor 7--10 smaller than the size
measured at 15\,GHz, which is close to the sizes usually measured for the
hot spots. Low frequency VLBI observations 
are needed to confirm the existence of such a compact component in 
the hot spot.

The lack of polarized emission at 320\,MHz for 1203$+$043 and a radio
structure which lacks a bright compact component rules out both
of the mechanisms for low frequency variability in this source.  
Consequently, we 
conclude that this source might be a spurious case of variability.
However,  1203$+$043 has an interesting structure at 320\,MHz.
It shows a pair of secondary lobes in a direction perpendicular to the
main source axis, making the object one of a few known `X'-shaped sources.
At present, only about ten sources are known to show such
morphology. They are believed to have both young and old lobes.  
These lobes may be supplied by jets whose direction has changed with
time.  A change in the orientation of the central engine due to 
precession has been suggested by Ekers et al.\ (1978) for 
NGC326 to account for its `X' shaped morphology.  Such a  model has
been applied successfully to 0828$+$32 by Klein et al.\ (1995) but
with the extra assumption that the length of the precessing beam
changes with time.   A merger between galaxies is thought to be the
cause of the precession.   However, Ulrich-Demoulin \& R\"onnback
(1996) have reported that optical images of 0828$+$32 do not show the 
signature of a recent major merger event.

However, the structure of 1203$+$043 looks peculiar when compared with other
'X' shaped sources.  For example, it has an asymmetric structure with
respect to the component C which is believed to be the core
(Fig.~7 and Fig.~6 in Mantovani et al. 1992). The long, bent jet is 
clearly 'one-sided' while,
generally speaking, the members of the class show two-sided jets
(at the available resolution). Moreover, the young lobes of the
'X' shaped sources are dominated by hot spots while here the jet emission
fades away from the core and the northern lobe contains only diffuse
emission without any bright component. This asymmetry is reflected in
the 320 MHz map where the region to the North-West is more extended
and brighter than the opposite side. 

\section {Conclusions}

We have conducted a program of multi-wavelength VLA observations
of the suspected low frequency variable sources 0235$-$197 and
1203+043.  Since 0235$-$197 is not polarized at
320\,MHz, its variability cannot be accounted for by instrumental 
polarization effects as in the case of 3C159.   0235$-$197 may
contain a low frequency component sufficiently compact and bright 
as required by the refractive scintillation model for low frequency 
variability.  Our observations have insufficient resolution to test
this suggestion; low frequency VLBI observations are required for this
purpose. However, this component would have to have extremely unusual 
properties among hot spots in radio sources.

In our high frequency
images of 1203+043 we have identified the core of the radio source;
its location indicates that the source has a large apparent asymmetry.  At
320\,MHz, this source shows no polarization.  However, it
does have an additional, steep-spectrum component at this frequency;
this previously-undetected component lies perpendicular to the main
axis and predominantly to one side.  However, the overall morphology
of 1203+043 at low frequencies seems similar to that of the `X'-shaped
sources like NGC326. From its morphology and component sizes,
we conclude that 1203+043 is likely not variable at low frequencies
and that its inclusion in such catalogs is spurious.

\acknowledgements{
The authors like to thank the referee, Dr. Steve Spangler for his
comments to the paper and Dr. Ian Browne for a critical reading of the
manuscript. 
FM thanks  Miller Goss, Assistant Director, NRAO, Socorro, for his hospitality
during period when part of the work was done. The National Radio
Astronomy Observatory is operated by Associated Universities Inc., under
cooperative agreement with the National Science Foundation; AIPS is 
NRAO's {\it Astronomical Image Processing System\/}.
}

%
% ---------- References -------------
%

%
%
\end{document}